\documentclass[10pt]{article}
\usepackage{newlfont}
\usepackage{amsmath}
\usepackage{times}
\usepackage[all]{xypic}
\usepackage{array,multirow,graphicx}
\usepackage[colorlinks]{hyperref}
\usepackage{authblk}
\ifpdf
    \graphicspath{{Fig/}{Fig/}{Fig/}}
\else
    \graphicspath{{Fig/}{Fig/}}
\fi
\newcommand{\sech}{\textmd{sech}}
\newcommand{\csch}{\textmd{csch}}
\numberwithin{equation}{section}
\begin{document}
\title{A Built-in Inflation in the $f(T)$-Cosmology}%
\author[1,2,3]{G.G.L. Nashed \thanks{nashed@bue.edu.eg}}
\author[1,2]{W. El Hanafy \thanks{waleed.elhanafy@bue.edu.eg}}
\affil[1]{\small \it Centre for theoretical physics at the British University in Egypt, 11837 - P.O. Box 43, Egypt.}
\affil[2]{\small \it Egyptian Relativity Group, Egypt.}
\affil[3]{\small \it Mathematics Department, Faculty of Science, Ain Shams University, Cairo, Egypt.}
\renewcommand\Authands{ and }
\date{\small Received: 29 April 2014 / Accepted: 24 September 2014}
\maketitle
\begin{abstract}
In the present work we derive an exact solution of an isotropic and homogeneous Universe governed by $f(T)$ gravity. We show how the torsion contribution to the FRW cosmology can provide a \textit{unique} origin for both early and late acceleration phases of the Universe. The three models ($k=0, \pm 1$) show a \textit{built-in} inflationary behavior at some early Universe time; they restore suitable conditions for the hot big bang nucleosynthesis to begin. Unlike the standard cosmology, we show that even if the Universe initially started with positive or negative sectional curvatures, the curvature density parameter enforces evolution to a flat Universe. The solution constrains the torsion scalar $T$ to be a constant function at all time $t$, for the three models. This eliminates the need for the dark energy (DE).  Moreover, when the continuity equation is assumed for the torsion fluid, we show that the flat and closed Universe models \textit{violate} the conservation principle, while the open one does not. The evolution of the effective equation of state (EoS) of the torsion fluid implies a peculiar trace from a quintessence-like DE to a phantom-like one crossing a matter and radiation EoS in between; then it asymptotically approaches a de Sitter fate.
\end{abstract}
\smallskip
Published version: Eur. Phys. J. C (2014) 74:3099. DOI 10.1140/epjc/s10052-014-3099-5\\
This article is distributed under the terms of the Creative Commons Attribution License. Funded by SCOAP$^{3}$ / License Version CC BY 4.0.
\section{Introduction}\label{S1}

Recent observational data suggest that our Universe is accelerating. Amongst the possible explanations for this phenomenon are modifications to gravitational theory \cite{NO7}. Cosmological constant, from an ideal fluid having different shapes of EoS with a negative pressure, a scalar field with quintessence-like or phantom-like behavior can also explain DE \cite{ENOSF}. It is not obvious what kind of DE is more suitable to explain the present epoch of the Universe. Observational data point to some type of DE having an EoS parameter which is close to $-1$, or even less than $-1$ (which is the phantom case). Modification of general relativity (GR) seems to be  quite attractive possibility to resolve the above mentioned problem. Modifications of the Hilbert-Einstein action through the introduction of general functions of the Ricci scalar $R$ have been extensively explored \cite{NO07}-\cite{M11}. These $f(R)$ gravity theories can be reformulated in terms of scalar field quintessence. Moreover, it has been shown  that when starting from $f(R)$ gravity, the phantom case in scalar tensor theory does not exist. However, when the conformal transformation becomes complex the phantom barrier is crossed, and therefore the resulting $f(R)$ function becomes complex. These cases are studied \cite{NOG} in more detail, in which, to avoid this handicap, a dark fluid was used to produce the phantom behavior such that  $f(R)$ function reconstructed from the scalar tensor theory continues to be real.

Initially the idea of teleparallelism theory has been proposed by Einstein in order to unify gravity and electromagnetism \cite{ Ea, OINC}. Later Einstein left the teleparallelism theory not because of its failure in the attempt of unification only, but also because of the vanishing of the curvature tensor of the Weitzenb\"{o}ck connection. But the non-vanishing torsion tensor aroused recently a great interest in astrophysical and cosmological applications in the so called $f(T)$ gravitational theory. The main motivations of such theory were:
\begin{itemize}
\item [(1)] GR can be viewed as a certain theory of teleparallelism; thus, it could be regarded  at least
as a different perspective  that could  lead to the same results \cite{TM}.

\item[(2)] In such a context, one can define energy and momentum tensors of the gravitational field which are  true
 tensors  under all general coordinate transformations but {\it not under local Lorentz transformation}.

\item [(3)]  This theory is interesting because it can be seen as gauge theories of the translation group (not the full Poincar\'{e} group); consequently, one may provide an alternative interpretation of GR \cite{Hw}-\cite{YE13}. Most recently Teleparallel Equivalent of General Relativity (TEGR) has been generalized to $f(T)$ theory, a theory of modified gravity formed in the same spirit as generalizing GR to $f(R)$ gravity \cite{BF,Le}. A main merit  of $f(T)$ gravity  theory is that its  gravitational field equation is of second order, the same as for GR, while it is of fourth order in metric $f(R)$ gravity. This merit makes the analysis of the cosmological expansion of the Universe in $f(T)$ gravity much easier than in $f(R)$ gravity. $f(T)$ gravity has gained significant attention in the literature with promising cosmological implications \cite{FF7}-\cite{RHSR}.
\end{itemize}

The main target of this work is to show how $f(T)$ gravity can be useful in explaining the flatness and acceleration at early and late phases of our Universe. For as is well known, current observations of the present Universe indicates that our Universe now is almost spatially flat. This leads one to exclude the closed and open Universe models. On the other hand the initial flat space assumption contradicts the presence of the strong gravitational field (i.e. the Riemann curvature) as it should! This contradiction might be explained as the flatness problem of the standard cosmology. Actually this problem has been overcome by the idea of an inflationary scenario during $\sim 10^{-36}-10^{-32}$ sec from the big bang. Lots of inflationary models have been proposed by using scalar fields. But to gain the benefits of both inflation and the standard cosmology the inflation should end at $\sim 10^{-32}$ sec from the big bang. This needs slow-roll conditions so that the inflationary Universe ends with a vacuum dominant epoch allowing the Universe to restore the big bang scenario. So the inflation can be considered as an add-on tool rather than a replacement of the big bang \cite{L03}. Until now there are no satisfactory reasons for the transition from inflation to big bang. Our trail here to treat these problems starts by diagnosing the core of the problem. We found that the curvature within the framework of the GR may lead to these conflicts, while introducing new qualities to the space-time, like torsion, might give a different insight of these problems.

The work is arranged as follows: In Section \ref{S2}, we describe the fundamentals of the $f(T)$ gravity theory. We next show the contribution of the torsion scalar field to the density and the pressure of the FRW models and necessary modifications in Section \ref{S3}. Also, we obtain a model dependent scale factor $R(t)$ and $f(T)$ as a solution of the continuity equation. In Section \ref{S4}, we investigate the cosmological behavior of the flat, closed and open Universe models due to $f(T)$ modifications. Moreover, we give the physical descriptions for the obtained results. In the flat Universe the teleparallel torsion scalar field $T$ and the $f(T)$ appear as constant functions and the later might replace the cosmological constant, the Universe shows an inflationary behavior as the scale factor $R(t) \propto e^{Ht}$, where the Hubble parameter $H$ is a constant. The flat Universe shows no evolution with time. Moreover, we investigate the closed Universe model which shows an inflationary behavior as well. In spite of the torsion scalar field $T$ appears as a constant function similar to the flat case, but the $f(T)$ of the closed Universe appears as a function of time. This allows the cosmological parameters to evolve. In particular the evolution of the curvature density parameter $\Omega_{k}$ shows a clear tendency to vanish at late time, which explains how the Universe can start with initial curvature; then it goes naturally to flat behavior. Combining the curvature density parameter within the total density parameter $\Omega_{\textmd{Tot}}$ in addition to the matter $\Omega_{m}$ and the torsion $\Omega_{T}$ density parameters gives a very restrictive range for the total density parameters $|\Omega_{\textmd{Tot}}-1| \leq 10^{-16}$ at some early time. This is a suitable value to begin the primordial nucleosynthesis epoch. The late accelerating expansion of the Universe is also recognized as the Hubble parameter $H > 0$ and the deceleration parameter $q \rightarrow -1$. Furthermore, the investigation of the open Universe shows a behavior similar to the closed model. So both closed and open models suggest a unique source for early and late acceleration phases of the Universe. While the open model, uniquely, implies a time dependent effective EoS of the torsion fluid. Its evolution starting initially with a quintessence-like energy to asymptotical de Sitter crosses radiation-, dust- and phantom-like energies. Section \ref{S5} is devoted to summarizing and concluding the results.
\section{ABC  of $f(T)$}\label{S2}

In the Weitzenb\"{o}ck space-time the fundamental field variables describing gravity are a quadruplet of parallel vector fields \cite{Wr}-\cite{NS} ${h_i}^\mu$, which we call the tetrad field characterized by
\begin{equation}\label{q1}
D_{\nu} {h_i}^\mu=\partial_{\nu}
{h_i}^\mu+{\Gamma^\mu}_{\lambda \nu} {h_i}^\lambda=0,
\end{equation}
where
${\Gamma^\mu}_{\lambda \nu}$ define the nonsymmetric affine connection
\begin{equation}\label{q2}
{\Gamma^\lambda}_{\mu \nu} \stackrel{\textmd{def.}}{=}
{h_i}^\lambda {h^i}_{\mu, \nu},
\end{equation}
with $h_{i \mu, \nu}=\partial_\nu h_{i \mu}$\footnote{space-time indices $\mu,\nu, \cdots$ and $SO$(3,1) indices $a, b, \cdots$ run from 0 to 3. Time and space indices are indicated by $\mu=0, i$, and $a=(0), (i)$.}. Equation (\ref{q1}) leads to the metricity  condition and the identically vanishing of curvature tensor defined by ${\Gamma^\lambda}_{\mu\nu}$, given by Equation (\ref{q2}). The metric tensor $g_{\mu \nu}$ is defined by
\begin{equation}\label{q3}
 g_{\mu \nu} \stackrel{\textmd{def.}}{=}  \eta_{i j} {h^i}_\mu {h^j}_\nu,
\end{equation}
with $\eta_{i j}=(+1,-1,-1,-1)$ is the metric of Minkowski space-time. We note that, associated with any tetrad field ${h_i}^\mu$ there  is a metric field defined uniquely by (\ref{q3}), while a given metric $g^{\mu \nu}$ does not determine the tetrad field completely; for any local Lorentz transformation of the tetrads ${h_i}^\mu$ leads to a new set of tetrads which also satisfy (\ref{q3}). Defining the torsion components and the contortion as
\begin{eqnarray}
\nonumber {T^\alpha}_{\mu \nu}  & \stackrel {\textmd{def.}}{=} &
{\Gamma^\alpha}_{\nu \mu}-{\Gamma^\alpha}_{\mu \nu} ={h_a}^\alpha
\left(\partial_\mu{h^a}_\nu-\partial_\nu{h^a}_\mu\right),\\
{K^{\mu \nu}}_\alpha  & \stackrel {\textmd{def.}}{=} &
-\frac{1}{2}\left({T^{\mu \nu}}_\alpha-{T^{\nu
\mu}}_\alpha-{T_\alpha}^{\mu \nu}\right), \label{q4}
\end{eqnarray}
where the contortion equals the difference between Weitzen\"{o}ck  and Levi-Civita connection, i.e., ${K^{\mu}}_{\nu \rho}= {\Gamma^\mu}_{\nu \rho }-\left \{_{\nu  \rho}^\mu\right\}$.  The tensor ${S_\alpha}^{\mu \nu}$ is defined as
\begin{equation}\label{q5}
{S_\alpha}^{\mu \nu}
\stackrel {\textmd{def.}}{=} \frac{1}{2}\left({K^{\mu
\nu}}_\alpha+\delta^\mu_\alpha{T^{\beta
\nu}}_\beta-\delta^\nu_\alpha{T^{\beta \mu}}_\beta\right),
\end{equation}
which is skew symmetric in the last two indices. The torsion scalar is defined as
\begin{equation}\label{q6}
T \stackrel {\textmd{def.}}{=} {T^\alpha}_{\mu \nu}
{S_\alpha}^{\mu \nu}.
\end{equation}
Similar to the $f(R)$ theory, one can define the action of $f(T )$ theory as
\begin{equation}\label{q7}
{\cal L}({h^a}_\mu, \Phi_A)=\int
d^4x h\left[\frac{1}{16\pi}f(T)+{\cal L}_{Matter}(\Phi_A)\right], \quad \textrm
{where} \quad h=\sqrt{-g}=det\left({h^a}_\mu\right),
\end{equation}
and we assumed the units in which $G = c = 1$  and  $\Phi_A$ are the matter fields.   Considering the action (\ref{q7}) as a function of the fields ${h^a}_\mu$ and putting  the variation of the function with respect to the field ${h^a}_\mu$ to be vanishing one can obtain the following equations of motion \cite{BF,CGSV}.
\begin{equation}\label{q8}
{S_\mu}^{\rho \nu} T_{,\rho} \
f(T)_{TT}+\left[h^{-1}{h^a}_\mu\partial_\rho\left(h{h_a}^\alpha
{S_\alpha}^{\rho \nu}\right)-{T^\alpha}_{\lambda \mu}{S_\alpha}^{\nu
\lambda}\right]f(T)_T-\frac{1}{4}\delta^\nu_\mu f(T)=-4\pi{{\cal T}^\nu}_\mu,
\end{equation}
where
$T_{,\rho}=\frac{\partial T}{\partial x^\rho}$, $f(T)_T=\frac{\partial f(T)}{\partial T}$, $f(T)_{TT}=\frac{\partial^2 f(T)}{\partial T^2}$ and ${{\cal T}^\nu}_\mu$ is the energy momentum tensor.

\section{Cosmological modifications of $f(T)$}\label{S3}

Recent cosmic observations support that the Universe is expanding with an acceleration. In this paper we attempted to apply the $f(T)$ field equations to the universe. In this cosmological model the Universe is taken as homogeneous and isotropic in space, which directly gives rise to the tetrad given by Robertson \cite{Rob}. This tetrad has the same metric as FRW metric; it can be written in spherical polar coordinate ($t$, $r$, $\theta$, $\phi$) as follows:
\begin{equation}\label{tetrad}
\left({h_{i}}^{\mu}\right)=\left(
  \begin{array}{cccc}
    1 & 0 & 0 & 0 \\
    0&\displaystyle\frac{L_1 \sin{\theta} \cos{\phi}}{4R(t)} & \displaystyle\displaystyle\frac{L_2 \cos{\theta} \cos{\phi}-4r\sqrt{k}\sin{\phi}}{4 r R(t)} & -\displaystyle\frac{L_2 \sin{\phi}+4 r \sqrt{k} \cos{\theta} \cos{\phi}}{4 r R(t)\sin{\theta}} \\[5pt]
    0&\displaystyle\frac{L_1 \sin{\theta} \sin{\phi}}{4 R(t)} & \displaystyle\frac{L_2 \cos{\theta} \sin{\phi}+4 r \sqrt{k}\cos{\phi}}{4 r R(t)} & \displaystyle\frac{L_2 \cos{\phi}-4 r \sqrt{k} \cos{\theta} \sin{\phi}}{4 r R(t)\sin{\theta}} \\[5pt]
    0&\displaystyle\frac{L_1 \cos{\theta}}{4 R(t)} & \displaystyle\frac{-L_2 \sin{\theta}}{4 r R(t)} & \displaystyle\frac{\sqrt{k}}{R(t)} \\[5pt]
  \end{array}
\right),
\end{equation}
where $R(t)$ is the \textit{scale factor}, $L_1=4+k r^{2}$ and $L_2=4-k r^{2}$. Substituting from the vierbein (\ref{tetrad}) into (\ref{q6}), we get the torsion scalar
\begin{equation}\label{Tscalar}
\begin{split}
   T=&\frac{6 k- 6 \dot{R}^2}{R^2},\\
    =&-6\left(H^2-\frac{k}{R^2},\right)\\
    =&-6H^2(1+\Omega_{k}),
\end{split}
\end{equation}
where $H(=\frac{\dot{R}}{R})$ is the \textit{Hubble} parameter and $\Omega_{k}(=\frac{-k}{R^2 H^2})$ is the \textit{curvature} energy density parameter.
The field equations (\ref{q8}) read
\begin{equation}\label{T00}
    \mathcal{T}_{0}^{~0}=\frac{-R^{2} f -12\dot{R}^{2} f_{T}}{4 R^{2}},
\end{equation}
\begin{equation}\label{T11}
\mathcal{T}_{1}^{~1}=\mathcal{T}_{2}^{~2}=\mathcal{T}_{3}^{~3}=\frac{4 k (R^{2} f_{T}+12 \dot{R}^{2} f_{TT})-R^{4} f -4R^2(R\ddot{R}+2\dot{R}^2)f_{T}+48\dot{R}^2(R\ddot{R}-\dot{R}^2)f_{TT}}{4 R^4},
\end{equation}
where the EoS is taken for a perfect fluid so that the energy-momentum tensor is ${\mathcal{T}^{\mu}}_{\nu}=\textmd{diag}(\rho,-p,-p,-p)$. Using (\ref{T00}), the perfect fluid density $\rho$ is given by
\begin{equation}\label{dens1}
    4 \pi \rho=\frac{R^{2} f +12\dot{R}^{2} f_{T}}{4 R^{2}},
\end{equation}
and using (\ref{T11}), the proper pressure $p$ of the perfect fluid is given by
\begin{equation}\label{press1}
    4 \pi p=\frac{4 k (R^{2} f_{T}+12 \dot{R}^{2} f_{TT})-R^{4} f -4R^2(R\ddot{R}+2\dot{R}^2)f_{T}+48\dot{R}^2(R\ddot{R}-\dot{R}^2)f_{TT}}{4 R^4}.
\end{equation}
Equations (\ref{dens1}) and (\ref{press1}) are the modified Friedmann equations in the $f(T)$-gravity in its generalized form. Then, the EoS parameter $\omega=\frac{p}{\rho}$ of the perfect fluid is given by
\begin{equation}\label{EoS1}
    \omega=-1+\frac{4k(R^2 f_{T}+12 \dot{R}^2 f_{TT})-4(R\ddot{R}-\dot{R}^2)[R^2 f_{T}-12 \dot{R}^2 f_{TT}]}{R^2(R^2 f+12 \dot{R}^2 f_{T})}.
\end{equation}
Considering the total energy density and pressure of the Universe behaves as the DE. Assuming the EoS of the DE, i.e.,  $p=-\rho$, we get from Eq. (\ref{EoS1}) an explicit form of $f(T)$  as:
\begin{equation}\label{fT1}
    f(T)=a+b~e^{\frac{1}{12}\left[\frac{R^2(\ddot{R} R-\dot{R}^2-k)}{\dot{R}^2(\ddot{R} R-\dot{R}^2+k)}\right]T},
\end{equation}
where $a$ and $b$ are constants of integration. The above equation indicates that there is a certain code relating  $f(T)$ to the scale factor $R(t)$ so that we should investigate possible compatibilities of these two functions. In the flat case, equation (\ref{fT1}) seems to be suitable to produce the de Sitter Universe (i.e. $T=-6 H^2= const.$) which implies that $\dot{T}=\ddot{R}R^2-\dot{R}^2=0$; see (\ref{Tscalar}). This produces unavoidable undetermined quantity in the $f(T)$ form. Even in non-flat cases, the successful exponential scale factor of the inflationary cosmology requires a constant torsion scalar, i.e. $\dot{T}=\ddot{R} R -\dot{R}^{2}+k=0$, again we get undefined quantity in the above $f(T)$ form. Later, in \S{\ref{S3.3}}, we will recall (\ref{fT1}) to show that enforcing the universal density to produce a DE, as we have just done, is not a functional code for the universe! So we do not advice using this treatment to approach the accelerating Universe.
\subsection{The FRW dynamical equations}\label{S3.1}

Let us assume  that the background is a non-viscous fluid. As we have mentioned, we can not enforce the total density and pressure to be a DE. Alternatively, we can study the torsion contribution to both $\rho$ and $p$ in the Friedmann dynamical equations by replacing $\rho \rightarrow \rho+\rho_{T}$ and $p \rightarrow p+p_{T}$, where $\rho$, $\rho_{T}$,  $p$ and $p_{T}$ are the matter density, the torsion density, the matter pressure and the torsion pressure respectively.
\begin{eqnarray}
  3\left(\frac{\dot{R}}{R}\right)^2=3H^2  &=& 8\pi \rho + 8\pi \rho_{T} - 3\frac{k}{R^2},\label{FRW1} \\
  3\left(\frac{\ddot{R}}{R}\right) =3 q H^2 &=& -4 \pi \left(\rho+3 p\right)-4 \pi \left(\rho_{T}+3 p_{T}\right), \label{FRW2}
\end{eqnarray}
where $q(=-\frac{R\ddot{R}}{\dot{R}^2})$ is the \textit{deceleration} parameter. In the above equation we take the general case of a non-vanishing pressure $p \neq 0$. It is clear that when $\rho_{T}=0$ and $p_{T}=0$ the above equations reduce to the usual Friedmann equations in GR. We take $\rho=\rho_{c}$ where $\rho_{c}$ is the critical density of the Universe when it is full of matter and spatially flat ($k=0$), then $\rho_{c}=\frac{3H^2}{8\pi}$. Substituting in equations (\ref{FRW1}) and (\ref{FRW2}) we get
\begin{eqnarray}
  1 &=& \Omega_{m}+\Omega_{T}+\Omega_{k},\label{FRW3} \\
  q &=& \frac{\left(\rho+3 p\right)/2}{3H^2/8 \pi}+\frac{\left(\rho_{T}+3 p_{T}\right)/2}{3H^2/8 \pi}, \label{FRW4}
\end{eqnarray}
where $\Omega_{m}=\frac{\rho}{\rho_{c}}=\frac{\rho}{3H^2/8\pi}$ represents the \textit{matter density} parameter and $\Omega_{T}=\frac{\rho_{T}}{\rho_{c}}=\frac{\rho_{T}}{3H^2/8\pi}$ represents the \textit{torsion density} parameter.

\subsection{The torsion contribution}\label{S3.2}

In order to obtain the torsion contribution $\rho_{T}$ and $p_{T}$, we rewrite equations (\ref{dens1}) and (\ref{press1}), in terms of the Hubble parameter, as below
\begin{equation}\label{dens2}
    4 \pi \rho=\frac{1}{4}(f+12 H^2 f_{T}).
\end{equation}
\begin{equation}\label{press2}
    4 \pi p=\frac{k}{R^2}(f_{T}+12H^2 f_{TT})-\left(\dot{H}+3H^2\right)f_{T}+12 \dot{H} H^2 f_{TT}-\frac{1}{4}f.
\end{equation}
Also, the EoS-parameter (\ref{EoS1}) can be rewritten as
\begin{equation}\label{q23}
    \omega=-1+\frac{4k(f_{T}+12 H^2 f_{TT})}{R^2(f+12 H^2 f_{T})}-\frac{4\dot{H}(f_{T}-12H^2 f_{TT})}{(f+12H^2 f_{T})}.
\end{equation}
Substituting the matter density that is obtained by the $f(T)$ field equation (\ref{dens2}) into the FRW dynamical equation (\ref{FRW1}), we get the torsion density
\begin{equation}\label{Tor_density}
    \rho_{T}=\frac{1}{8 \pi}\left(3H^2-f/2-6H^2 f_{T}+\frac{3k}{R^2}\right).
\end{equation}
The above equation can be written in the form
\begin{equation*}
    \frac{\rho_{T}}{3H^2/8\pi}=1-\left[\frac{f}{6 H^2}+2 f_{T}\right]+\frac{k}{H^2 R^2},
\end{equation*}
so that the torsion density parameter is
\begin{equation}\label{tor_dens_par}
    \Omega_{T}=1-\left[\frac{f}{6 H^2}+2 f_{T}\right]-\Omega_{k},
\end{equation}
comparing the above equation to equation (\ref{FRW3}) we get the modified matter density parameter as
\begin{equation}\label{matt_density_par}
    \Omega_{m}=\frac{f}{6H^2}+2 f_{T}.
\end{equation}
Similarly we substitute from (\ref{dens2}), (\ref{press2}) and (\ref{Tor_density}) into (\ref{FRW2}) we get
\begin{equation}\label{Tor_press}
    p_{T}=\frac{-1}{8 \pi}\left[\frac{k}{R^2}(1+2f_{T}+24H^2 f_{TT})+2\dot{H}+3H^2-f/2-2(\dot{H}+3H^2)f_{T}+24\dot{H}H^2 f_{TT}\right].
\end{equation}
The EoS parameter due to the torsion contribution is thus
\begin{equation}\label{Tor_EoS_par}
    \omega_{T}=\frac{p_{T}}{\rho_{T}}=-1+2/3\frac{(1-f_{T}+12H^2f_{TT})\dot{H}-(1-f_{T}-12H^2 f_{TT})k/R^2}{f/6-(1+2f_{T})H^2-k/R^2}.
\end{equation}
It is clear that $\omega_{T}=-1$ for the case of flat Universe ($k=0$ and $\dot{H}=0$), c.f. \cite{KA12}. The torsion contributes to the FRW model in a way similar to the cosmological constant.

\subsection{A generalized $R(t)$ and $f(T)$ as an ordered pair}\label{S3.3}

The scale factor $R(t)$ plays the key role in the Universe evolution and composition. Most of the cosmological applications leaves the scale factor to be chosen! In this section, we aim to get a generalized form for a \textit{model dependent} $f(T)$ and $R(t)$. In this case some solutions will be rejected due to incompatibility. This can be done as follows, we substitute the matter density (\ref{dens1}) and pressure (\ref{press1}) into the continuity equation
\begin{equation}\label{Cont_eqn}
    \dot{\rho}+3(\rho+p)\frac{\dot{R}^2}{R^2}=0,
\end{equation}
the continuity equation reads
\begin{equation}\label{Cont_eqn_DE}
    \dot{R}(\ddot{R}R-\dot{R}^2+k)(12f_{TT}\dot{R}^2+f_{T}R^2)=0.
\end{equation}
The solution of the above differential equation has many possible cases: We exclude the case of $R(t)$ is a constant as it gives a steady Universe. We interested to examine the case of the vanishing of the first and second brackets simultaneously. So we first take $\ddot{R}R-\dot{R}^2+k=0$, this constrains the torsion scalar to be a constant function with respect to time. By solving for the scale factor $R(t)$ we get
\begin{equation}\label{scale-factor}
    R(t)=\frac{c_{1}}{2}\left[\frac{e^{\frac{2(t+c_{2})}{c_{1}}}-k}{e^{\frac{(t+c_{2})}{c_{1}}}}\right],
\end{equation}
where $c_1$ and $c_2$ are constants of integration. The above equation provides an exponentially expanding Universe which is suitable for the inflationary scenario at the early time. We check the compatibility of (\ref{fT1}) and (\ref{scale-factor}), as mentioned in \S {\ref{S3}}, by substituting from (\ref{scale-factor}) into (\ref{fT1}) we get a forbidden case as the total energy density and pressure of the Universe cannot be a DE, as expected!

We next examine the vanishing of the second bracket of (\ref{Cont_eqn_DE}) so that $12f_{TT}\dot{R}^2+f_{T}R^2=0$, by substituting from (\ref{scale-factor}) and solving for $f(T)$ we get
\begin{equation}\label{fT}
    f(T)=c_{3}+c_{4}e^{\frac{-1}{12}\left[\frac{c_{1}(e^{\frac{-2(t+c_2)}{c_1}}-k)}{e^{\frac{-2(t+c_2)}{c_1}}+k}\right]^{2}T},
\end{equation}
where $c_3$ and $c_4$ are constants of integration. Equations (\ref{scale-factor}) and (\ref{fT}) verify the continuity equation (\ref{Cont_eqn}). One can easily show that the above $f(T)$ form is suitable to describe the acceleration of the late Universe. But, it is valid for flat and non-flat Universe models. Also, it covers perfectly equation (\ref{fT1}) without undetermined quantities in the $f(T)$ form. Combining the compatible solutions (\ref{scale-factor}) and (\ref{fT}) provides a consistent treatment to study both early and late Universe acceleration in flat and non-flat Universe models. So the obtained solution represents a generalized $f(T)$-gravity and a generalized scale factor $R(t)$ suitable for this study.

Also, It is worth to re-mention that the substitution from (\ref{scale-factor}) into (\ref{Tscalar}), implies a generalized behavior for the torsion scalar $T$ to be a constant function of the time $t$. So we may conclude that the torsion contribution to the energy density may not vary with time and dose not affected by the expansion of the Universe which is very similar to the behavior of the DE. We next examine the obtained solution in different world models.

\section{World Models}\label{S4}
One of the benefits of the obtained solution that it is a generalized $f(T)$ and $R(t)$ formula valid for the three world models, the spatially flat Universe ($k=0$), the pseudo sphere, open, Universe ($k=-1$) and the sphere, closed, Universe ($k=+1$). This enables us to examine the behavior of the DE and its effects on the cosmological parameters in these different models as follows.
\subsection{Flat Universe}
In the case of spatially-flat FRW universe, $k=0$, the scale factor (\ref{scale-factor}) becomes
\begin{equation}\label{sc-factk=0}
    R(t)=\frac{c_{1}}{2}e^\frac{t+c_{2}}{c_{1}},
\end{equation}
and (\ref{fT}) will be
\begin{equation}\label{fT_k=0}
    f(T)=c_3+c_4 e^{-\frac{1}{12}c_1^2 T}.
\end{equation}
It is convenient to reexpress some quantities in terms of the scale factor (\ref{sc-factk=0}): the Hubble parameter $H$; the torsion scalar $T$, (\ref{Tscalar}); the Hubble parameter change $\dot{H}$ (the dot represents the derivative with respect to time) and the declaration parameter $q$, respectively are
\begin{equation}\label{Hubble_k=0}
    H=1/c_{1},~~~ T= -6/{c_{1}}^{2},~~ ~\dot{H}=0,~~~q=-1.
\end{equation}
It is clear that $T=-6H^2$, one may use $T$ and $H$ exchangeably. Also, it is clear that the scale factor (\ref{sc-factk=0}) and the Hubble parameter (\ref{Hubble_k=0}) show an inflationary behavior of the Universe, where $H$ is a constant and $R(t) \propto e^{H t}$. One can easily conclude that the torsion scalar plays the role of the cosmological constant during the inflation period. We next evaluate the critical density, using (\ref{Hubble_k=0}), for flat space is
\begin{equation}\label{crit_dens_k=0}
    \rho_c=\frac{3/c_1^2}{8 \pi},
\end{equation}
the matter density (\ref{dens2}) and pressure (\ref{press2}) read
\begin{equation}\label{dens_k=0}
    \rho=\frac{c_3+c_4 \sqrt{e}}{16 \pi}=-p,
\end{equation}
the torsion density (\ref{Tor_density}) and pressure (\ref{Tor_press}) are
\begin{equation}\label{Tor_dens_k=0}
    \rho_T=\frac{6-(c_3+c_4 \sqrt{e}) c_1^2}{16 \pi c_1^2}=-p_T.
\end{equation}
One can easily find that the total density is at its critical value exactly $\rho_c=\rho+\rho_T$. Also, it should be mentioned that assuming the torsion fluid fulfills the continuity equation gives a case similar to the steady state cosmology, where the $\dot{\rho}=0$ and $\dot{\rho}_T=0$. The EoS parameter for both matter and torsion are
\begin{equation}\label{EoS_Par_k=0}
    \omega=-1,~ \omega_{T}=-1
\end{equation}
the curvature density parameter for the flat Universe $\Omega_k=0$, while the matter density parameter (\ref{matt_density_par}) is
\begin{equation}\label{matt_dens_par_k=0}
    \Omega_m=\frac{c_1^2}{6}(c_3+c_4 \sqrt{e}),
\end{equation}
and the torsion energy parameter (\ref{tor_dens_par}) is
\begin{equation}\label{Tor_dens_par_k=0}
    \Omega_T=1-\frac{c_1^2}{6}(c_3+c_4 \sqrt{e}).
\end{equation}
The above cosmological parameters show that the scale factor (\ref{sc-factk=0}) growths exponentially with time. But the Universe constituents do not change with time. This does not allow the Universe to evolute. However, the Universe shows an accelerated expansion. The equations (\ref{dens_k=0}), (\ref{Tor_dens_k=0}) and the continuity (\ref{Cont_eqn}) lead to the conclusion that the total density has a constant value, nevertheless the universe is expanding! This leads directly to a violation of the conservation principle of energy. In the following two Sections, we are going to examine similar cases in both the closed and open Universes.
\subsection{Sphere, closed, Universe}
In the case of the closed FRW Universe, $k=+1$, the scale factor (\ref{scale-factor}) becomes
\begin{equation}\label{sc_factk+1}
    R(t)=-c_{1}\sinh\left(\frac{t+c_{2}}{c_{1}}\right),
\end{equation}
and (\ref{fT}) will be
\begin{equation}\label{fT_k+1}
   f(T)=c_{3}+c_{4}e^{-\frac{c_1^2}{12} \tanh^2 \left(\frac{t+c_2}{c_1}\right)T}.
\end{equation}
Using the above values for the scale factor and the torsion function we get the following cosmological parameters: The Hubble parameter
\begin{equation}\label{Hubblek+1}
    H=\frac{1}{c_1}\coth\left(\frac{t+c_2}{c_1}\right),
\end{equation}
\begin{equation}\label{Hdotk+1}
    \dot{H}=\frac{1}{c_1^2} \csch^{2}\left(\frac{t+c_2}{c_1}\right).
\end{equation}
The Hubble parameter $H$ appears in the closed Universe as a function of time not a constant as given in the flat case, but keeping the same exponential behavior of the scale factor with time as the flat Universe. We find this case is more suitable to describe the evolution of the constituents of the Universe. Another cosmological parameter which is related to the Universe evolution is the deceleration parameter, this parameter appears for the closed Universe as a function of time as
\begin{equation}\label{deck+1}
    q=-\tanh^2\left(\frac{t+c_2}{c_1}\right),
\end{equation}
In order to show the cosmological behavior, the deceleration parameter (\ref{deck+1}) versus redshift $z=\frac{R_{0}}{R}-1$, where $R_{0}$ is the scale factor at the present time, is plotted in Figure 1(a). The graph shows that the deceleration parameter $q \rightarrow 0$ as $z \rightarrow \infty$, then $q \rightarrow -1$ as $z \rightarrow 0$ at late Universe. The plot shows that the accelerating phases of the closed Universe from early to late time. Also, the graph shows that the deceleration parameter is $-1$ when the torsion scalar field is dominant. The curvature density parameter for the closed Universe is
\begin{equation}\label{curv-densk+1}
    \Omega_k=-\sech^2\left(\frac{t+c_2}{c_1}\right),
\end{equation}
In the standard Cosmology it is will known that if there is a slight deviation from the flat Universe, it goes to be more and more curved one very quickly. The the curvature density parameter in the closed Universe model initially chosen to produce a closed Universe. The cosmological parameter $\Omega_k$, given by (\ref{curv-densk+1}), is plotted versus the redshift $z$ in Figure 1(b). Unlike the standard cosmology the evolution of the curvature density parameter turns the Universe to be a flat one. This has a great interest in solving the flatness problem of the big bang cosmology. The torsion scalar (\ref{Tscalar}) becomes
\begin{figure}
\begin{center}
\includegraphics[scale=.3]{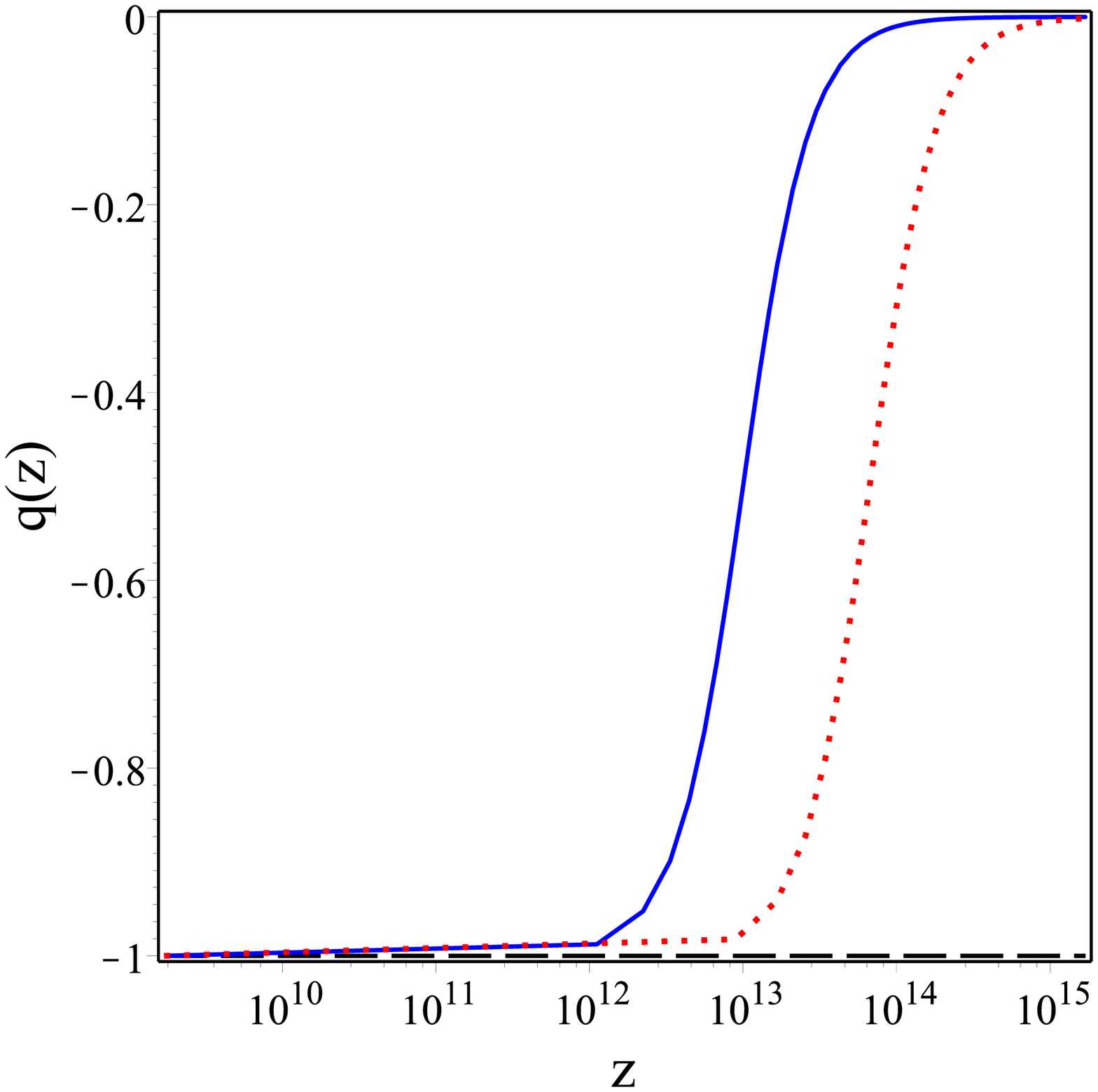}\label{Fig1a}\hspace{1cm}
\includegraphics[scale=.3]{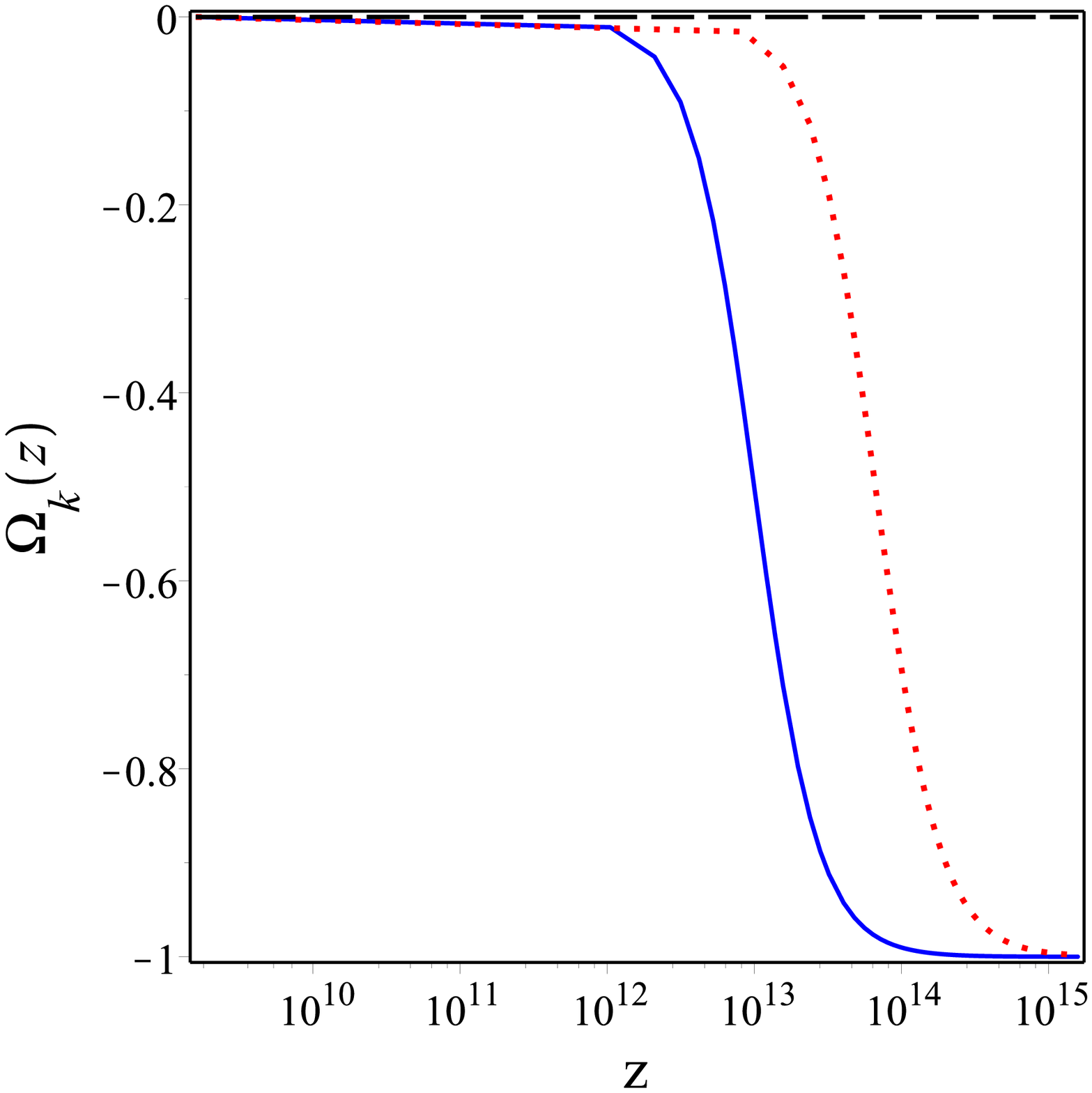}\label{Fig1b}
\caption{(a) The evolution of the deceleration parameter $q$ versus the redshift $z$. (b) The plot shows the evolution of the curvature densities parameters $\Omega_k$ versus the redshift. Here the dot and solid lines are for the constant $c_1=10^{-13}$ and $10^{-14}$, respectively. While the dash line is for $c_1\rightarrow 0$, alternatively when the torsion scalar field is dominant.}
\end{center}
\end{figure}
\begin{equation}\label{Tsck+1}
    T=-\frac{6}{c_1^2}.
\end{equation}
One should note that in spite of the Hubble and the curvature density parameters are functions of time they combine in a way to rule out the evolution of the torsion scalar field with the expansion. Also, it is clear that the torsion scalar field is dominant $T \rightarrow \infty$ as $c_{1} \rightarrow 0$. The critical density, for a closed universe, is generalized to be a function of time
\begin{equation}\label{crit_densk+1}
    \rho_c=\frac{3/c_1^2}{8 \pi} \coth^2\left(\frac{t+c_2}{c_1}\right),
\end{equation}
the matter density (\ref{dens2}) and pressure (\ref{press2}) are
\begin{equation}\label{densk+1}
    \rho=\frac{1}{16 \pi}\left[c_3+c_4 e^{\frac{1}{2}\tanh^2\left(\frac{t+c_2}{c_1}\right)}\right]=-p,
\end{equation}
while the torsion density (\ref{Tor_density}) and pressure (\ref{Tor_press}) read
\begin{equation}\label{Tor_densk+1}
    \rho_T=\frac{-1}{16 \pi}\left[c_3+c_4 e^{\frac{1}{2}\tanh^2\left(\frac{t+c_2}{c_1}\right)}+\frac{6}{c_1^2}\left(1-2 \coth^2\left(\frac{t+c_2}{c_1}\right) \right)\right]=-p_T.
\end{equation}
For this case of a closed Universe the matter and the torsion densities are no longer constants. But their equations of state evolute in a way similar to the flat Universe. Assuming the torsion fluid fulfills the continuity equation, one can easily find that $\dot{\rho}=0$ and $\dot{\rho}_T=0$. Again as in the flat Universe, we conclude that the closed Universe also violates the conservation principle of energy, and the EoS parameter for both matter and torsion give
\begin{equation}\label{EoS_par_k+1}
    \omega=-1,~\omega_{T}=-1,
\end{equation}
the matter density parameter (\ref{matt_density_par}) for the closed Universe reads
\begin{equation}\label{mat_dens_parak+1}
    \Omega_m=\frac{c_1^2}{6} \left[\frac{c_3+c_4 e^{\frac{1}{2}\tanh^2\left(\frac{t+c_2}{c_1}\right)}}{\coth^2\left(\frac{t+c_2}{c_1}\right)}\right],
\end{equation}
while the torsion density parameter (\ref{tor_dens_par}) becomes
\begin{equation}\label{Tor_den_parak+1}
    \Omega_T=1+\sech^2\left(\frac{t+c_2}{c_1}\right)-\frac{c_1^2}{6} \left[\frac{c_3+c_4 e^{\frac{1}{2}\tanh^2\left(\frac{t+c_2}{c_1}\right)}}{\coth^2\left(\frac{t+c_2}{c_1}\right)}\right].
\end{equation}
\begin{figure}
\begin{center}
\includegraphics[scale=.3]{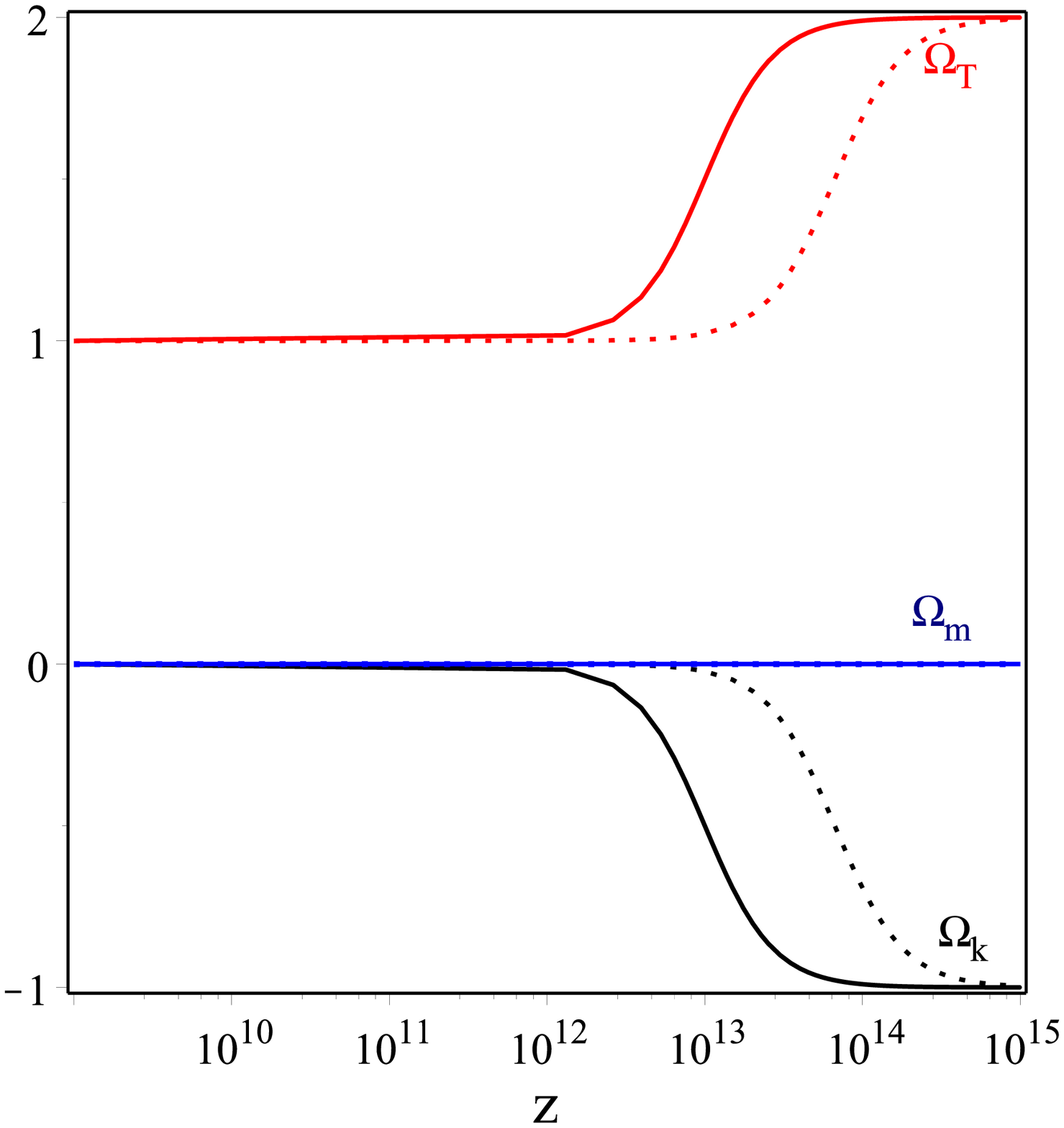}\label{Fig2a}\hspace{1cm}
\includegraphics[scale=.38]{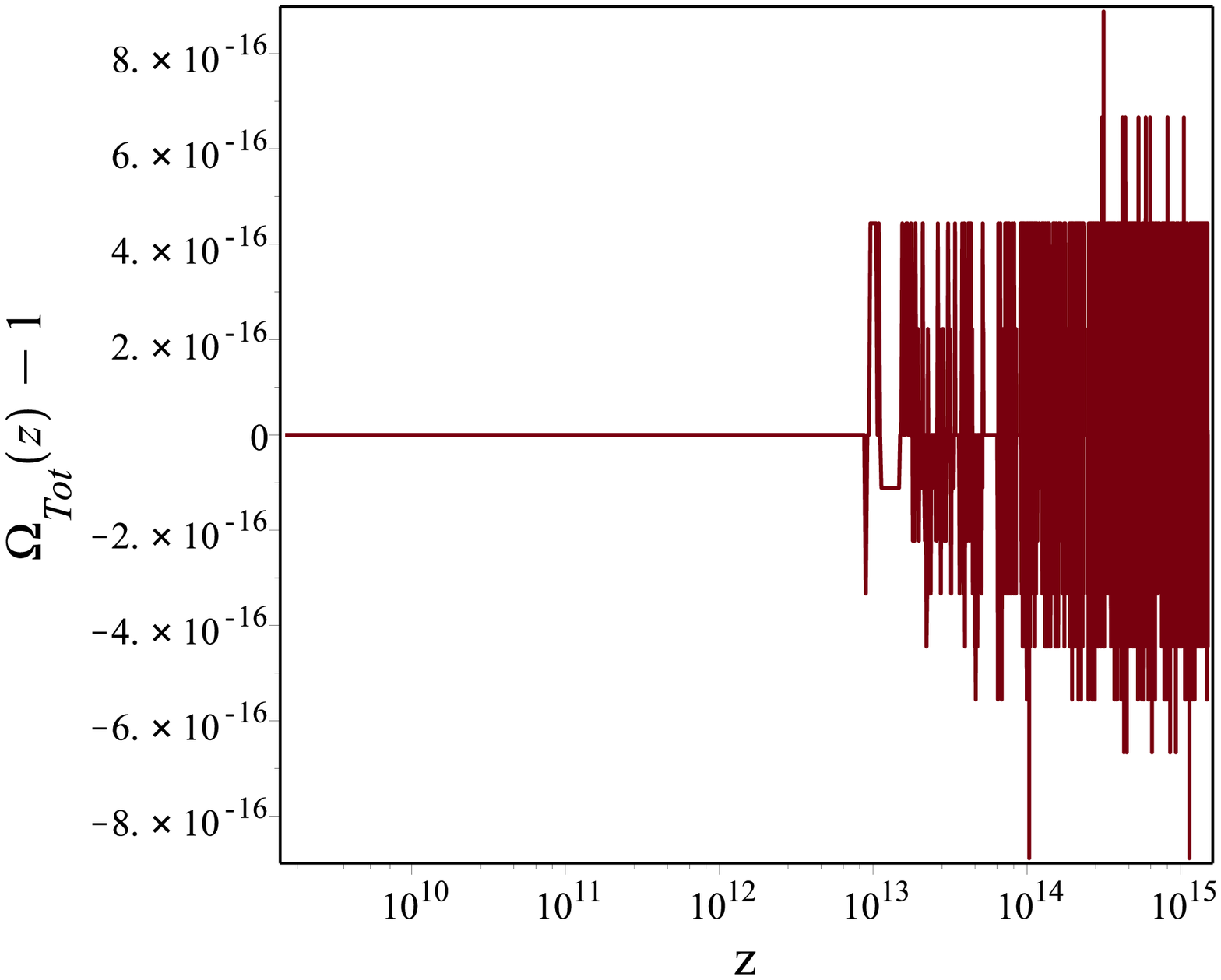}
\caption{(a) The plot shows the evolution of the density parameters $\Omega_k$, $\Omega_m$ and $\Omega_T$ versus the redshift $z$. The black, blue and red colours are for the curvature, matter and torsion density parameters, respectively. (b) The plot shows the evolution of the total density parameter $\Omega_{\textmd{Tot}} := \Omega_k+\Omega_m+\Omega_T$ versus the redshift. The dot and solid are correspond to the value of $c_1$ as in Figure 1.}
\end{center}
\end{figure}
The cosmological parameters in equations (\ref{curv-densk+1}), (\ref{mat_dens_parak+1}) and (\ref{Tor_den_parak+1}) are plotted versus the redshift $z$ in Figure 2(a) to provide information about the evolution of the cosmos components during the expansion for the closed Universe model. In spite of all the Universe compositions vary with time, we found a rapid change at early Universe, then it converges all compositions to act as a steady behavior of the flat Universe at late Universe. This leads to investigate the global behavior of the Universe compositions. Thus we define the total density parameter $\Omega_{\textmd{Tot}}:= \Omega_{m}+\Omega_{T}+\Omega_{k}$, where it includes the curvature one. According to the FRW dynamical equation (\ref{FRW1}), the total density parameter $\Omega_{\textmd{Tot}}$ initially equals to $1$. The early variation of the densities parameters in Figure 2(a) is reflected on the total density parameter, see Figure 2(b). The plot shows very high frequency variations which is explained by recognizing the rapid, but smooth, variation of the densities parameters at early time, then it turns back to $1$ at late Universe when the parameters become steady. The inflationary behavior of (\ref{sc_factk+1}) combined with the violent variations shown in Fig 2(b) of an amplitude of $|\Omega_{\textmd{Tot}}-1| \leq 10^{-16}$ restores the most outstanding success of the Hot big bang, the nucleosynthesis.

Also, the obtained closed Universe model shows a behavior different from the standard cosmology. It is well known that when the Universe is slightly shifted from the flat case it goes further away to be more curved which is inconsistent with the present observation. This leads to assume an initial flat Universe model. Here we show that the Universe might start initially with a positive curvature then it turns to a flat Universe behavior. This reopens the closed Universe model for more investigations.

In addition, the calculations of the cosmological parameters for the closed Universe model (\ref{densk+1}) and (\ref{Tor_densk+1}) show that a case similar to the flat Universe. Where the total density of the Universe is constant, $\dot{\rho}=0$ and $\dot{\rho}_T=0$, while the Universe expands! Again, we get a violation to the energy conservation principle.
\subsection{Pseudo sphere, open, Universe}
In the case of the open FRW Universe, $k=-1$, the scale factor (\ref{scale-factor}) becomes
\begin{equation}\label{sc_factk-1}
    R(t)=c_{1}\cosh\left(\frac{t+c_{2}}{c_{1}}\right),
\end{equation}
and (\ref{fT}) will be
\begin{equation}\label{fT_k-1}
    f(T)=c_{3}+c_{4}e^{-\frac{c_1^2}{12} \coth^2 \left(\frac{t+c_2}{c_1}\right)T}.
\end{equation}
Using the above values of the scale factor and the torsion function we get: The Hubble parameter
\begin{equation}\label{Hubblek-1}
    H=\frac{1}{c_1}\tanh\left(\frac{t+c_2}{c_1}\right),
\end{equation}
the Hubble parameter appears as a function of time whose gradual change in time as
\begin{equation}\label{Hdotk-1}
    \dot{H}=\frac{1}{c_1^2} \sech^{2}\left(\frac{t+c_2}{c_1}\right),
\end{equation}
Also here in the open Universe case we got an exponential scale factor but a varying Hubble parameter. This case is more suitable to find the evolution of the Universe. The deceleration parameter will be
\begin{equation}\label{deck-1}
    q=-\coth^2\left(\frac{t+c_2}{c_1}\right),
\end{equation}
we plot the deceleration parameter versus the redshift $z$ in Figure 3(a). The evolution of the deceleration parameter versus the redshift $z$ shows a possible deceleration epoch when $q > 0$ before going to be negative allowing accelerated expansion of the open Universe. Also, the curvature density parameter is given by
\begin{equation}\label{curv-densk-1}
    \Omega_k=\csch^2\left(\frac{t+c_2}{c_1}\right).
\end{equation}
The evolution of the curvature density parameter is plotted versus the redshift $z$ in Figure 3(b). The plot shows that the curvature density parameter started initially with an arbitrary value then it converges naturally to the flat case, which agrees with the present Universe observations. This encourages to reconsider the curved open Universe model.
\begin{figure}
\begin{center}
\includegraphics[scale=.3]{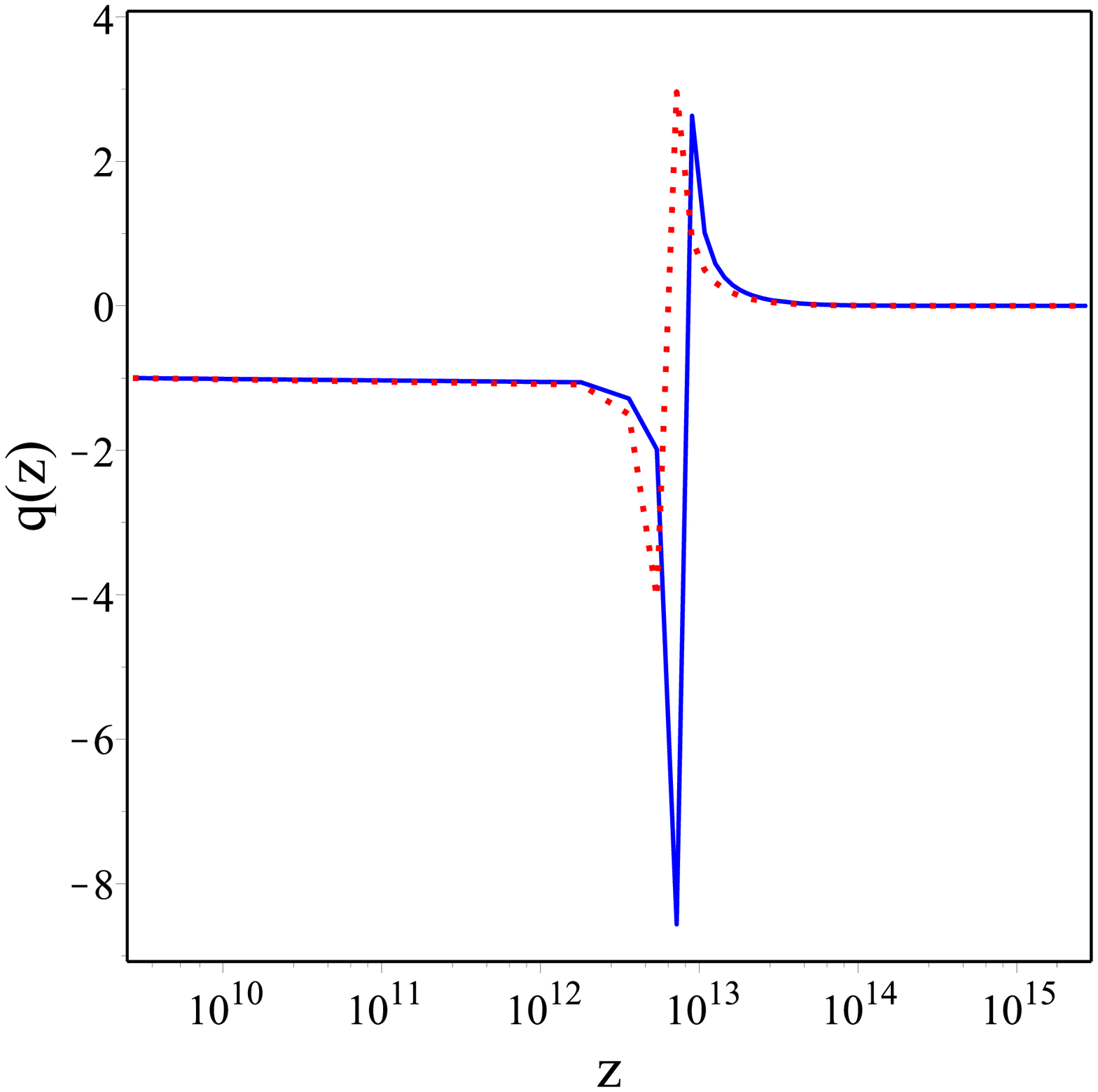}\label{Fig3a}\hspace{1cm}
\includegraphics[scale=.3]{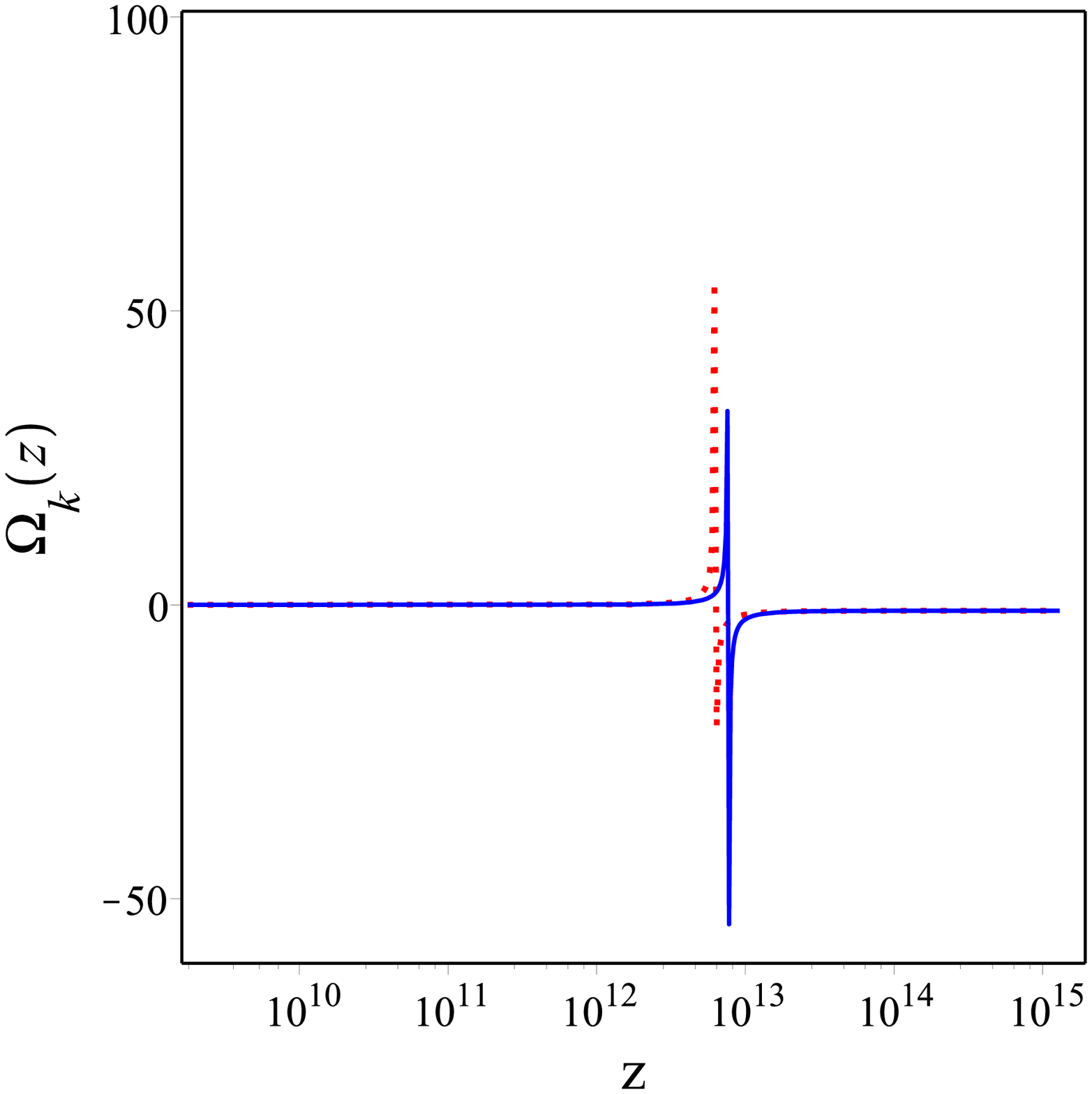}\label{Fig3b}
\caption{(a) The deceleration parameter versus the redshift $z$. Here the solid and dot lines are for the constant $c_1=1.3 \times 10^{-13}$ and $1.6 \times 10^{-13}$, respectively. (b) The plot shows the evolution of the curvature density parameter versus the redshift, the solid and dot are correspond to the value of $c_1$ as in (a).}
\end{center}
\end{figure}
We next evaluate the torsion scalar field (\ref{Tscalar}) in the open Universe, we get
\begin{equation}\label{Tsck-1}
    T=-\frac{6}{c_1^2},
\end{equation}
it should be mentioned here that, in the case of an open Universe, we found that the scale factor (\ref{sc_factk-1}) and the Hubble parameter (\ref{Hubblek-1}) combine in a way to cancel out the effect of the time on the evolution of the torsion scalar field. The the critical density for an open Universe is given as a function of time as
\begin{equation}\label{crit_densk-1}
    \rho_c=\frac{3/c_1^2}{8 \pi} \tanh^2\left(\frac{t+c_2}{c_1}\right),
\end{equation}
while the matter density (\ref{dens2}) and pressure (\ref{press2}) are
\begin{equation}\label{densk-1}
    \rho=\frac{1}{16 \pi}\left[c_3+c_4 e^{\frac{1}{2}\coth^2\left(\frac{t+c_2}{c_1}\right)}\right]=-p,
\end{equation}
Combining the above result with the continuity equation (\ref{Cont_eqn}) gives, $\dot{\rho} = 0$, a constant value of the matter density with the expansion. This implies a continuous creation of matter. The EoS for the matter gives
\begin{equation}\label{Eos_para_k-1}
    \omega=-1.
\end{equation}
Also, the torsion density (\ref{Tor_density}) for the open Universe reads
\begin{equation}\label{Tor_densk-1}
    \rho_T=\frac{-1}{16 \pi}\left[c_3+c_4 e^{\frac{1}{2}\coth^2\left(\frac{t+c_2}{c_1}\right)}+\frac{6}{c_1^2} \left(1-2 \tanh^2\left(\frac{t+c_2}{c_1}\right)\right)\right],
\end{equation}
and the torsion pressure (\ref{Tor_press}) becomes
\begin{equation}\label{Tor_pressk-1}
    p_T=\frac{1}{16 \pi}\left[c_3+c_4 e^{\frac{1}{2}\coth^2\left(\frac{t+c_2}{c_1}\right)}-\frac{2}{c_1^2}\left(1+2\tanh^2\left(\frac{t+c_2}{c_1}\right)\right)\right],
\end{equation}
Assuming that the torsion fluid fulfills the continuity equation, this implies that $\dot{\rho}_T \neq 0$. The evolution of the torsion fluid prevents a violation of the conservation principle.\\

It is clear that the torsion density and pressure (\ref{Tor_densk-1}) and (\ref{Tor_pressk-1}) implies that $p_{T} \neq -\rho_{T}$. The EoS Parameter of the torsion (\ref{Tor_EoS_par}) appears as a function of time
\begin{equation}\label{Tor_Eos_parak-1}
    \omega_T=-1+\frac{8~ \sech^2\left(\frac{t+c_2}{c_1}\right)}{c_1^2\left[c_3+c_4 e^{\frac{1}{2}\coth^2\left(\frac{t+c_2}{c_1}\right)}+\frac{6}{c_1^2}\left(1-2 \tanh^2\left(\frac{t+c_2}{c_1}\right)\right)\right]}.
\end{equation}

We recognize that the open Universe case, uniquely, gives a dynamical behavior of the EoS of the torsion fluid. The evolution of the EoS parameter, (\ref{Tor_Eos_parak-1}) shows an initial quintessence-like DE, crossing $\omega_{T}=0$ dust like epoch to a radiation one at $\omega_{T} \sim \frac{1}{3}$ then it turns back crossing $\omega_{T}=0$ very quickly to cross $\omega_{T}=-1$ implying that a phantom-like DE ($\omega_{T}<-1$), then it asymptotically approaches a de Sitter fate, see Fig. 4. It is well known that the density of the phantom-like dark torsion fluid $\rho_{T} \propto R(t)^n$, where $n$ is positive, which implies an increasing of the density as the Universe expands. The phantom energy epoch might be created as a result of the curvature density parameter decay in order to preserve the energy conservation principle.
\begin{figure}
\begin{center}
\includegraphics[scale=.33]{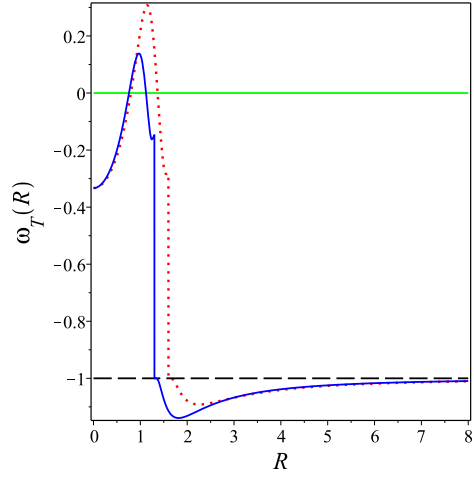}\label{Fig4}
\caption{The evolution of the EoS of the torsion fluid (\ref{Tor_Eos_parak-1}) versus the scale factor. The torsion fluid acts as a \textit{quintessence}-like DE ($-1<\omega_{T}<0$) at high redshift, while it asymptotically approaches a de sitter fate in the future crossing $\omega_{T}=0$ and $\omega_{T}=-1$ in between. The constants $c_3=2$ and $c_4=-7$ mainly control the amplitude of the torsion EoS parameter crossing $\omega_{T}=0$ and $\omega_{T}=-1$. The solid and dot lines are for the constant $c_1=1.3$ and $1.6$, respectively. The dash line is for $c_1 \rightarrow 0$ when the torsion scalar field is dominant which acts as the cosmological constant.}
\end{center}
\end{figure}
We next write the matter density parameter (\ref{matt_density_par}) for the open Universe as
\begin{equation}\label{mat_den_parask-1}
    \Omega_m=\frac{c_1^2}{6} \left[\frac{c_3+c_4 e^{\frac{1}{2}\coth^2\left(\frac{t+c_2}{c_1}\right)}}{\tanh^2\left(\frac{t+c_2}{c_1}\right)}\right],
\end{equation}
while the torsion density parameter (\ref{tor_dens_par}) will be
\begin{equation}\label{Tor_den_parak-1}
    \Omega_T=1-\csch^2\left(\frac{t+c_2}{c_1}\right)-\frac{c_1^2}{6} \left[\frac{c_3+c_4 e^{\frac{1}{2}\coth^2\left(\frac{t+c_2}{c_1}\right)}}{\tanh^2\left(\frac{t+c_2}{c_1}\right)}\right].
\end{equation}
The open Universe model provides information about the evolution of the Universe compositions during the expansion. Equations (\ref{curv-densk-1}), (\ref{mat_den_parask-1}) and (\ref{Tor_den_parak-1}) have been plotted versus the redshift $z$ in Fig 5(a). However, the Universe compositions vary with time very quickly, they combine later in a way to give a flat Universe behavior. The investigation of the global behavior of the Universe compositions shows that the total density parameter $\Omega_{\textmd{Tot}}$ is extremely closed to $1$. Figure 5(b) shows that a very restrictive variation range of the total density parameter $|\Omega_{\textmd{Tot}}-1| \leq 10^{-15}$ at early Universe, which is similar to the closed Universe case but slightly less. Then it turns to $1$ at some late Universe time. However, the Universe shows an inflationary behavior, (\ref{sc_factk-1}), it restores the critical value of $\Omega_{\textmd{Tot}}$ for the nucleosynthesis to begin. We must mention here that the open Universe model is the most accurate model, in the present work, as the nucleosynthesis epoch is from $\sim 1 \rightarrow 200$ seconds, while similar case of the closed Universe takes much longer time.

In addition, the calculations of the cosmological parameters for the open Universe model (\ref{densk-1}) and (\ref{Tor_densk-1}) show that a case similar to the flat Universe for the matter content where the matter density of the Universe is constant, $\dot{\rho}=0$. Again, by assuming that the torsion fluid fulfills the continuity equation, we find a behavior different from the flat or the closed models. Since, the torsion density is not constant, $\dot{\rho}_T \neq 0$, while the Universe expands! We get a unique behavior of the open Universe model prevents the violation of the energy conservation principle. This leads to the conclusion that the torsion density might decay reproducing a matter density as the Universe expands. Moreover, the open Universe uniquely implies an initial quintessence-like and later phantom-like energy and a de Sitter in the future. For the above mentioned reasons we find that the open Universe model is the most accurate and consistent model in the present work. We summarize the evaluated cosmological parameters of the three models in the next subsection.
\begin{figure}
\begin{center}
\includegraphics[scale=.32]{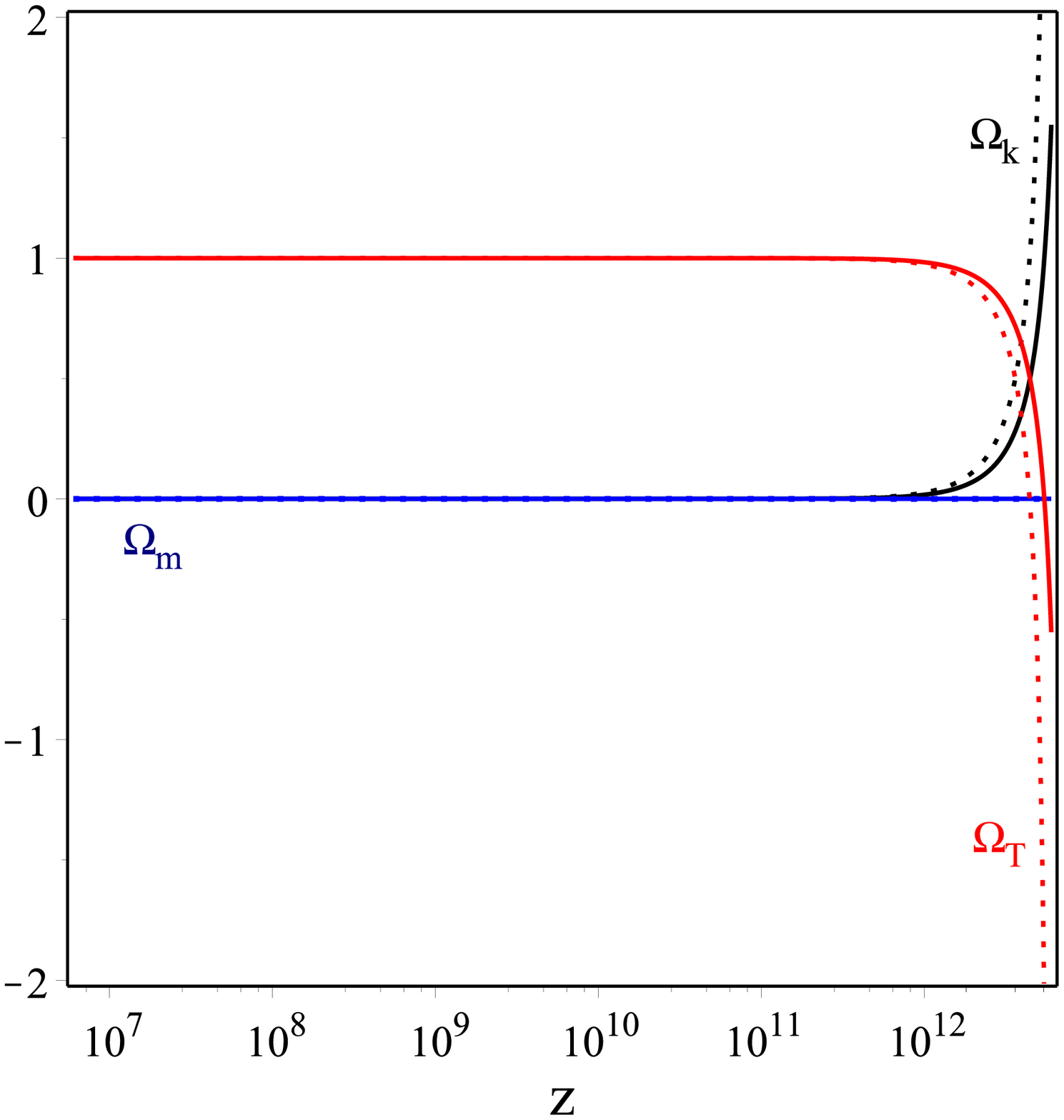}\label{Fig5a}\hspace{1cm}
\includegraphics[scale=.4]{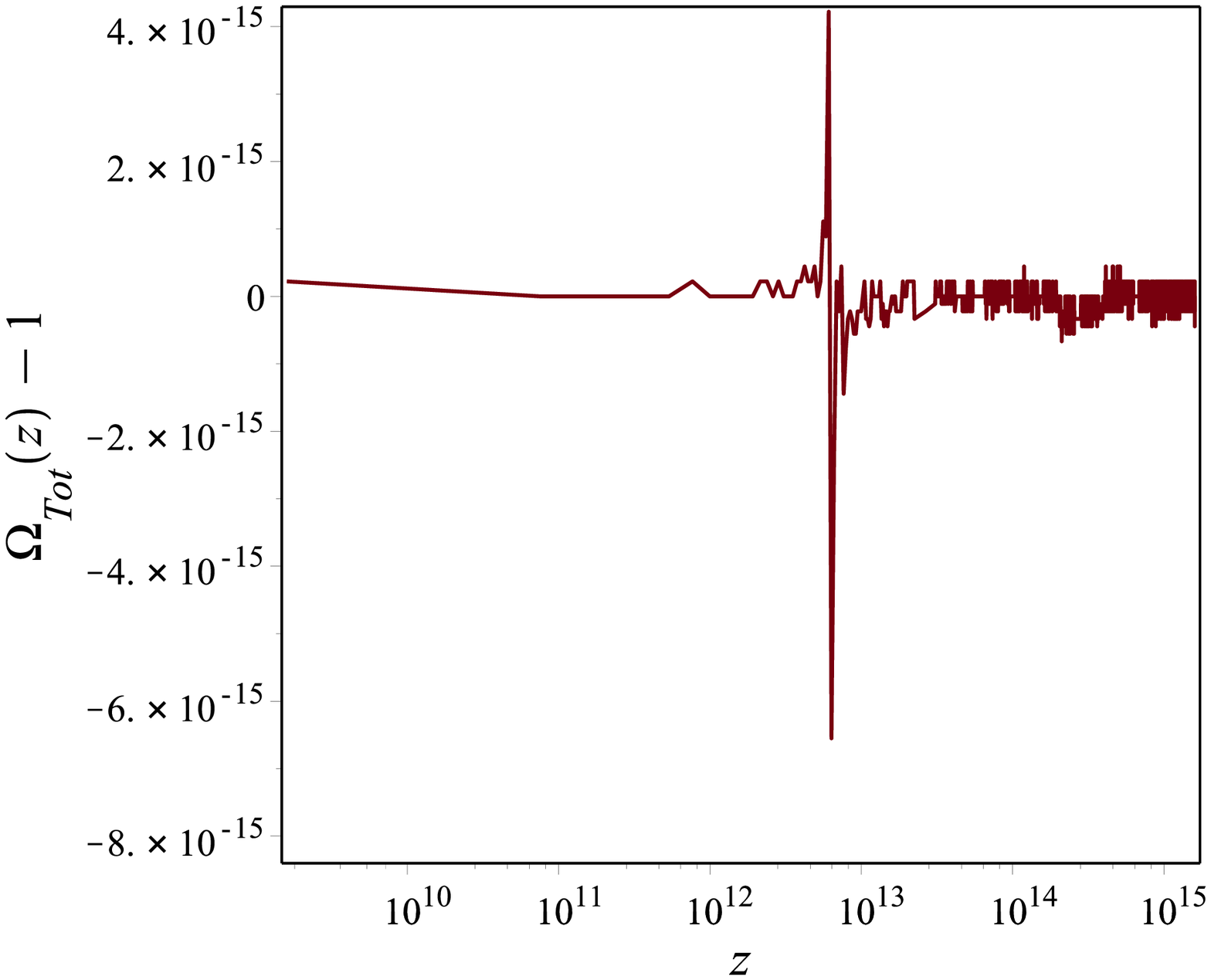}\label{Fig5b}
\caption{(a) The plot shows the evolution of the density parameters $\Omega_k$, $\Omega_m$ and $\Omega_T$ versus the redshift $z$. The black, blue and red colours are for the curvature, matter and torsion density parameters, respectively. The solid and dot are correspond to the value of $c_1=1.3 \times 10^{-13}$ and $1.6 \times 10^{-13}$, respectively. (b) The plot shows the evolution of the total density parameter $\Omega_{\textmd{Tot}} := \Omega_k+\Omega_m+\Omega_T$ versus the redshift for $c_1=1.6 \times 10^{-13}$. }
\end{center}
\end{figure}
%
\subsection{Cosmological parameters summary in the three world models}
We summarize the calculated cosmological parameters for the three world models $k=0, \pm 1$ and list it in Table 1. These values are useful to discuss the standard problems of cosmology, i.e. the particle horizon, the flatness and the singularity problems.

We may split these cosmological parameters to two different sets: the first is to describe the composition of the Universe, which contains three parameters $\Omega_{m}$, $\Omega_{k}$ and $\Omega_{T}$. The second set is to describe the expansion of the Universe, which contains two parameters $H$ and $q$.
\begin{table}\label{Table1}
\begin{center}
 \caption{Summary of The Cosmological Parameters.}
\begin{tabular}{|l|c|c||c|c|}
\hline
& \multicolumn{2}{c||}{evolution}   & \multicolumn{2}{c|}{composition}   \\
& \multicolumn{2}{c||}{cosmological parameters}   & \multicolumn{2}{c|}{cosmological parameters}   \\
\hline
 $k=-1$ &        & $c_{1}\cosh\left(\frac{t+c_{2}}{c_{1}}\right)$              &            & $\csch^2\left(\frac{t+c_2}{c_1}\right)$ \\[5pt]
 $k=0 $ & $R(t)$ & $\frac{1}{2}c_{1}\exp\left(\frac{t+c_{2}}{c_{1}}\right)$    & $\Omega_k$ & $0$ \\[5pt]
 $k=+1$ &        & $-c_{1}\sinh\left(\frac{t+c_{2}}{c_{1}}\right)$              &            & $-\sech^2\left(\frac{t+c_2}{c_1}\right)$ \\[5pt]
\hline
 $k=-1$ &        & $\tanh\left(\frac{t+c_2}{c_1}\right)/c_{1}$ &            & $\frac{c_1^2}{6} \left[\frac{c_3+c_4 e^{\frac{1}{2}\coth^2\left(\frac{t+c_2}{c_1}\right)}}{\tanh^2\left(\frac{t+c_2}{c_1}\right)}\right]$ \\[5pt]
 $k=0 $ & $H$  & $1/c_{1}$                                   & $\Omega_m$ &  $\frac{c_1^2}{6}(c_3+c_4 \sqrt{e})$\\[5pt]
 $k=+1$ &        & $\coth\left(\frac{t+c_2}{c_1}\right)/c_{1}$ &            & $\frac{c_1^2}{6} \left[\frac{c_3+c_4 e^{\frac{1}{2}\tanh^2\left(\frac{t+c_2}{c_1}\right)}}{\coth^2\left(\frac{t+c_2}{c_1}\right)}\right]$ \\[5pt]
\hline
 $k=-1$ &        & $-\coth^{2}\left(\frac{t+c_{2}}{c_{1}}\right)$ &            & $1-\csch^2\left(\frac{t+c_2}{c_1}\right)-\frac{c_1^2}{6} \left[\frac{c_3+c_4 e^{\frac{1}{2}\coth^2\left(\frac{t+c_2}{c_1}\right)}}{\tanh^2\left(\frac{t+c_2}{c_1}\right)}\right]$ \\[5pt]
 $k=0 $ & $q$    & $-1$                                           & $\Omega_T$ & $1-\frac{c_1^2}{6}(c_3+c_4 \sqrt{e})$  \\[5pt]
 $k=+1$ &        & $-\tanh^{2}\left(\frac{t+c_{2}}{c_{1}}\right)$ &            & $1+\sech^2\left(\frac{t+c_2}{c_1}\right)-\frac{c_1^2}{6} \left[\frac{c_3+c_4 e^{\frac{1}{2}\tanh^2\left(\frac{t+c_2}{c_1}\right)}}{\coth^2\left(\frac{t+c_2}{c_1}\right)}\right]$\\[5pt]
\hline
\end{tabular}
\end{center}
\end{table}
\section{Concluding Remarks}\label{S5}
\begin{itemize}
  \item In this work we have evaluated the matter density and pressure of the $f(T)$ field equations. We modified the FRW models due to the torsion contribution by replacing $\rho \rightarrow \rho+\rho_T$ and $p \rightarrow p+p_T$. Most of the cosmological models choose the scale factor $R(t)$ independent of the model. In this work we have got a model dependent $R(t)$ and $f(T)$ as order pairs, when applying the continuity equation to the Universe matter assuming that the torsion scalar and time are independent variables. The obtained solutions allow us to study the three world models, i.e. $k=0, \pm 1$. The calculations show that the torsion scalar (\ref{Tscalar}) can be written as a combination of the Hubble parameter $H$ and the curvature density parameter $\Omega_k$. These two parameters always combine keeping the torsion scalar a constant at all time $t$.
  \item The study of the \textit{flat} Universe model produces an inflationary cosmological model $R(t) \propto e^{Ht}$, $H=const$. But the Universe's compositions have no evolution where the matter density is constant during the expansion $\dot{\rho}=0$. Assuming the continuity equation for the torsion fluid leads to a constant torsion density, $\dot{\rho}_{T}=0$, during the expansion. This gives a steady state Universe. The total density of the Universe is equivalent to a constant Universe critical density. Then we conclude that the flat Universe model violates the conservation principle.
  \item The cosmological parameters for the \textit{closed} Universe model are found as functions of time. These parameters show a quick evolution at some early Universe, then they show a steady behavior at later time. Although the Universe in the closed model is chosen to be curved initially, the Universe's composition enforces the Universe to be flat at some late time as $\Omega_{\textmd{Tot}} \rightarrow 1$ and $\Omega_k \rightarrow 0$. Assuming the continuity equation for the torsion fluid implies a case similar to the flat model.
  \item In the case of the \textit{open} Universe model we have found a quick evolution of the cosmological parameters at some early time. The Universe in this model has been chosen to be initially curved, while the evolution of the cosmological parameters turns the Universe to be flat at some later time. The calculations show that the evolution of the open Universe prevents the violation of the conservation principle. This makes the open Universe model the most acceptable one.
  \item The inflationary Universe has been started as a speculative idea to solve some problems of the big bang cosmology. The inflation has been considered as an add-on extra tool to the standard big bang during some very early Universe. In this model, we get a built-in inflationary behavior at early time and then the model enables the big bang to be restored naturally.
  \item In the standard big bang cosmology it is known that the Universe becomes more and more curved very quickly, if it has been chosen to be initially curved, i.e. $\Omega_{\textmd{Tot}}$ diverges away from the unity. But the current cosmological observations show that our present Universe is almost flat. This requires a flat Universe initial condition. In our model, unlike the standard cosmology, we found that even if the Universe has started with an initial curvature, the evolution of $\Omega_{\textmd{Tot}}$ converges to unity. This tells that the Universe in the case of $k = \pm 1$ models is enforced to be flat. This solves many of the hot big bang cosmology problems. The closed Universe model shows that an extremely restrictive range for the total density parameter $|\Omega_{\textmd{Tot}}-1| \leq 10^{-16}$ at early Universe time, which is required for the nucleosynthesis epoch to begin and restore the big bang scenario. The open Universe shows almost the same restrictive range but much shorter interval of time. The result agrees with the BBN period ($\sim 1-200$ sec.) which again supports the open Universe model. See Figures 3(b) and 5(b).
  \item In the open model we have found that the teleparallel torsion fluid explains both early and late cosmic acceleration. This eliminates the need for the DE; in addition, it does not address the cosmological constant problem. Also, the use of the torsion scalar instead of the cosmological constant gives a conservative Universe. In addition, the torsion contribution gives a built-in inflationary behavior at a very early time; then the evolution of the total density parameter $\Omega_{\textmd{Tot}}$ shows good agreement with later stages. Moreover, the open Universe converges to a flat one, which agrees perfectly with the current observations. Furthermore, the evolution of the torsion fluid EoS, Fig. 4, shows a peculiar dynamical behavior during different phases of the cosmic expansion. There are many other details of these models that need further investigations. In particular, one would be interested in the torsion density and pressure in the open Universe model and their possible justifications as regards quantum cosmology.
\end{itemize}
\subsection*{ACKNOWLEDGMENTS}
This work is partially supported by the Egyptian Ministry of Scientific Research under project No. 24-2-12.
\subsection*{Open Access}
This article is distributed under the terms of the Creative Commons Attribution License which permits any use, distribution, and reproduction in any medium, provided the original author(s) and the source are credited.\\
Funded by SCOAP$^{3}$ / License Version CC BY 4.0.

\begin{thebibliography}{99}

\bibitem{NO7} S. Nojiri and  S.D. Odintsov, {\it Int. J. Geom. Meth. Mod. Phys}, {\bf 4} (2007), 115.

\bibitem{ENOSF} E. Elizalde, S. Nojiri, S. D. Odintsov, D. Saez-Gomez and V. Faraoni, {\it Phys. Rev.}
{\bf D77} (2008), 106005.

\bibitem{NO07} S. Nojiri and S. D. Odintsov,  {\it J. Phys. Conf. Ser} {\bf 66} (2007), 012005 (2007).

\bibitem{NO6}  S. Nojiri and S. D. Odintsov, {\it Phys. Rev.} {\bf D74} (2006), 086005.

\bibitem{CNOT} S. Capozziello, S. Nojiri, S. D. Odintsov and A. Troisi, {\it Phys. Lett.} {\bf B 639} (2006), 135.

\bibitem{NOG}  S. Nojiri and S. D. Odintsov and D. S\'aez-G\'omez, {\it  Phys. Lett.}  {\bf B681} (2009) 74.

\bibitem{CEOT}  G. Cognola, E. Elizalde, S. D. Odintsov, P. Tretyakov and S. Zerbini, {\it Phys. Rev.} {\bf D79} (2009), 044001.

\bibitem{NOG}  E.  Elizalde and D. S\'aez-G\'omez {\it Phys. Rev.} {\bf D80} (2009), 044030.

\bibitem{M11} R. Myrzakulov, {\it Eur. Phys. J. } {\bf C71} (2011), 1752.

\bibitem{Ea} A. Einstein, {\it Sitzungsber. Preuss. Akad. Wiss. Phys. Math. Kl.}, (1928) 217 (1930) 401.

\bibitem{OINC} Y. C. Ong, K. Izumi,. J. M. Nester, P. Chen, {\it arXiv:1303.0993v1}

\bibitem{TM} T. Ort\'{i}n {\rm "Gravity and Strings"} Cambridge
University Press (2004), P. 166.

\bibitem{Hw} F. W. Hehl, {\it in Proceedings of the 6th School of Cosmology and
Gravitation on Spin, Torsion, Rotation and Supergravity}, Erice,
1979, edited by P. G. Bergmann and V. de Sabbata (Plenum, New
York, 1980).

\bibitem{HMMN95} F. W. Hehl, J. D. McCrea, E. W. Mielke and Y. Ne'eman, {\it Phys. Rep.} {\bf 258}, 1 (1995).

\bibitem{HS} K. Hayashi, {\it Phys. Lett.} {\bf 69B}, 441 (1977).

\bibitem{HS79} K. Hayashi and T. Shirafuji. {\it Phys. Rev.} {\bf D19}, 3524 (1979).

\bibitem{HS81} K. Hayashi and T. Shirafuji. {\it Phys. Rev.} {\bf D24}, 3312 (1981).

\bibitem{BV88} M. Blagojevi$\acute{c}$ and M. Vasili$\acute{c}$  {\it Class. Quant. Grav.} {\bf 5} (1988), 1241.

\bibitem{K00} T. Kawai, {\it Phys. Rev. } {\bf D62} (2000), 104014.

\bibitem{KST00} T. Kawai, K. Shibata and I. Tanaka, {\it Prog. Theor. Phys.} {\bf 104} (2000), 505.

\bibitem{YS07} N.L. Youssef and A.M. Sid-Ahmed, {\it Rep. Math. Phys.} {\bf 60} (2007), pg. 39--53.

\bibitem{YS08} N.L. Youssef and A.M. Sid-Ahmed, {\it Int. Jour. Geom. Meth. Mod. Phys.} {\bf 5} (2008), pg. 1109.

\bibitem{WYS10} M.I. Wanas, N.L. Youssef and A.M. Sid-Ahmed, {\it Class. Quant. Grav.} {\bf 27} (2010), pg. 045005.

\bibitem{YE13} N.L. Youssef and W.A. Elsayed, {\it Rep. Math. Phys.} {\bf 72} (2013), pg. 1--23.

\bibitem{BF} G. R. Bengochea and R. Ferraro, {\it Phys. Rev.} {\bf D 79} (2009), 124019.

\bibitem{Le} E. V. Linder, {\it Phys. Rev.} {\bf D 81} (2010), 127301.

\bibitem{FF7} R. Ferraro and F. Fiorini,  {\it Phys.
Rev.} {\bf D 75} (2007), 084031.

\bibitem{FF8} R. Ferraro and F. Fiorini,  {\it Phys.
Rev.} {\bf D 78} (2008), 124019.

\bibitem{CDDS} H. Chen, J. B. Dent, S. Dutta and E. N. Saridakis, {\it  Phys.
Rev.} {\bf D 83} (2011), 023508.

\bibitem{WY11} P. Wu, H. W. Yu, {\it Eur. Phys. J.} {\bf C71} (2011), 1552.

\bibitem{DDS} J. B. Dent, S. Dutta, E. N. Saridakis, {\it  JCAP}  {\bf 1101} (2011), 009.

\bibitem{ZH11}  R. Zheng, Q. -G. Huang, {\it JCAP} {\bf 1103} (2011), 002.

\bibitem{BGLL} K. Bamba, C. -Q. Geng, C. -C. Lee, L. -W. Luo, {\it JCAP} {\bf 1101} (2011), 021.

\bibitem{Yr11} R. -J. Yang, {\it Europhys. Lett.} {\bf 93} (2011), 60001.

\bibitem{WY10} P. Wu, H. W. Yu, {\it  Phys. Lett.} {\bf B693} (2010), 415.

\bibitem{Bg11} G. R. Bengochea, {\it Phys. Lett.} {\bf B695} (2011), 405.

\bibitem{WY101} P. Wu, H. W. Yu, {\it Phys. Lett.} {\bf B692} (2010), 176.

\bibitem{ZLGZ} Y. Zhang, H. Li, Y. Gong, Z. -H. Zhu,  {\it JCAP} {\bf 1107} (2011), 015.

\bibitem{CCDDS} Y. -F. Cai, S. -H. Chen, J. B. Dent, S. Dutta, E. N. Saridakis, {\it Class. Quant. Grav.} {\bf 28} (2011), 2150011.

\bibitem{CD11} S. Chattopadhyay, U. Debnath, {\it theories, Int. J. Mod. Phys.}  {\bf D20} (2011), 1135.

\bibitem{SR11} M. Sharif, S. Rani, {\it Mod. Phys. Lett.} {\bf A26} (2011), 1657.

\bibitem{WMQ} H. Wei, X. -P. Ma, H. -Y. Qi, {\it  Phys. Lett.} {\bf B703} (2011), 74.

\bibitem{BMT} C. G. Boehmer, A. Mussa, N. Tamanini, {\it Class. Quant. Grav.} {\bf 28} (2011), 245020.

\bibitem{WQM} H. Wei, H. -Y. Qi, X. -P. Ma, {\it  Eur. Phys. J.} {\bf C 72} (2012), 2117.

\bibitem{CCFR}  S. Capozziello, V. F. Cardone, H. Farajollahi, A. Ravanpak, {\it Phys. Rev.} {\bf D 84} (2011), 043527.

\bibitem{DRH} M. H. Daouda, M. E. Rodrigues, M. J. S. Houndjo, {\it Eur. Phys. J.} {\bf C 72} (2012), 1890.

\bibitem{BG} K. Bamba, C. -Q. Geng,  {\bf JCAP} {\bf 1111} (2011), 008.

\bibitem{GLSW} C. -Q. Geng, C. -C. Lee, E. N. Saridakis, Y. -P. Wu, {\it Phys. Lett.} {\bf B704} (2011), 384.

\bibitem{Wh12} H. Wei, {\it  Phys. Lett.} {\bf B 712} (2012), 430.

\bibitem{GLS} C. -Q. Geng, C. -C. Lee, E. N. Saridakis, {\it JCAP} {\bf 1201} (2012), 002.

\bibitem{BHL}  C. G. B\"oehmer, T. Harko and F. S. N. Lobo, {\it  Phys. Rev.} {\bf D 85} (2012), 044033.

\bibitem{AD12} K. Atazadeh and F. Darabi, {\it  Eur. Phys. J.} {\bf C 72}
(2012), 2016.

\bibitem{JMSM} M. Jamil, D. Momeni, N. S. Serikbayev and R. Myrzakulov, {\it  Astrophys. Space Sci.} {\bf 339} (2012), 37.

\bibitem{FRW} H. Farajollahi, A. Ravanpak and P. Wu, {\it  Astrophys. Space Sci.} {\bf 338} (2012), 23.

\bibitem{KA12} K. Karami and A. Abdolmaleki, {\it JCAP} {\bf 1204} (2012), 007.

\bibitem{YLZL} J. Yang, Y. -L. Li, Y. Zhong and Y. Li, {\it Phys. Rev.} {\bf  D 85} (2012), 084033.

\bibitem{XSL} C. Xu, E. N. Saridakis and G. Leon, {\it JCAP} {\bf 1207} (2012), 005.

\bibitem{BMNO} K. Bamba, R. Myrzakulov, S. \'{i}. Nojiri and S. D. Odintsov, {\it  Phys. Rev.} {\bf D 85} (2012),
104036.

\bibitem{CGSV} S. Capozziello, P. A. Gonz\'{a}lez, E. N. Saridakis, Y. V\'{a}squez, {\it JHEP} {\bf 1302} (2013), 039.

\bibitem{SH12} M. R. Setare and M. J. S. Houndjo, {\it Can. J. Phys.} {\bf 91} (2013) 260.

\bibitem{LWY} D. Liu, P. Wu and H. Yu, {\it Int. J. Mod. Phys.} {\bf D21} (2012),1250074.

\bibitem{IS12}  L. Iorio and E. N. Saridakis, {\it Mon. Not. Roy. Astron. Soc.} {\bf 427} (2012), 1555.

\bibitem{DWM} H. Dong, Y. -b. Wang and X. -h. Meng,{\it Eur. Phys. J.} {\bf C 72} (2012), 2002.

\bibitem{DRH12} M. H. Daouda, M. E. Rodrigues and M. J. S. Houndjo, {\it [arXiv:1205.0565]}.

\bibitem{BCNO} K. Bamba, S. Capozziello, S. Nojiri, S. D. Odintsov, {\it  Astrophysics and Space Science} {\bf 342} (2012), 155.

\bibitem{BSN} A. Behboodi, S. Akhshabi and K. Nozari, {\it Phys. Lett.} {\bf B 718} (2012), 30.

\bibitem{BF12} A. Banijamali and B. Fazlpour, {\it Astrophys Space Sci} {\bf 342} (2012), 229.

\bibitem{Mr12} R. Myrzakulov, {\it  The European Physical Journal} {\bf C72} (2012), 2203.

\bibitem{LR12} D. Liu and M. J. Reboucas, {\it   Phys.Rev.} {\bf D86} (2012), 083515.

\bibitem{W09} M.I. Wanas, {\it Amer. Inst. Phys. Conf. Ser.} {\bf 1115} (2009), pg. 218--223.

\bibitem{W12} M.I. Wanas, {\it Adv. High Ener. Phys.} {\bf 2012} (2012), Article ID 752613, 10 pages.

\bibitem{WH14} M.I. Wanas and H.A. Hassan, {\it Int. Jour. Theoret. Phys.} {\bf 53} (2014), pg. 3901--3909.

\bibitem{GC12} R. Ghosh and S. Chattopadhyay, {\it Eur. Phys. J. Plus} (2013), 128.

\bibitem{RHSR} M. E. Rodrigues, M. J. S. Houndjo, D. Saez-Gomez and F. Rahaman, {\it Phys. Rev.} {\bf D 86} (2012), 104059.

\bibitem{L03} A. Liddle, {\it An Introduction to Modern Cosmology}, Second Edition, May (2003).

\bibitem{Wr} R. Weitzenb\"ock, {\it Invariance Theorie, Nordhoff, Gronin-gen, 1923}.

\bibitem{CGSV} S. Capozziello, P. A. Gonzalez, E. N. Saridakis  and Y. Vasquez, {\it JHEP} {\bf 1302} (2013), 039.

\bibitem{MDTC} J. W. Maluf, J. F. da Rocha-neto, T. M. L. Toribio and K. H.
Castello-Branco,  {\it Phys.\ Rev.\ }{\bf D65} (2002), 124001.

\bibitem{NS} G.G.L. Nashed and T. Shirafuji, {\it Int. J. Mod. Phys.} {\bf D16} No. 1, (2007) 65.

\bibitem{Rob} H.P. Robertson, {\it Ann. Math.} {\bf 33} (1932), 496.
\end{thebibliography}

\end{document}